\documentclass[12pt, draftcls, onecolumn]{IEEEtran}
\usepackage{graphicx}
\usepackage{setspace}
\usepackage{float}
\usepackage{multicol}
\usepackage{grfext}
\usepackage {flushend, cuted}

\begin{document}

\title{Wireless Power Transfer: Survey and Roadmap}
\author{
Xiaolin~Mou,~\IEEEmembership{Student Member,~IEEE} and 
Hongjian~Sun,~\IEEEmembership{Member,~IEEE}
 \thanks{The paper is accepted to be published in Proceedings of IEEE VTC 2015 Spring Workshop on ICT4SG. This is a preprint version. Copyright (c) 2015 IEEE. Personal use of this material is permitted. However, permission to use this material for any other purposes must be obtained from the IEEE by sending a request to pubs-permissions@ieee.org.}
 \thanks{X. Mou and H. Sun are with the School of Engineering and Computing Sciences, Durham University, Durham, UK. (e-mail: xiaolin.mou@durham.ac.uk; hongjian@ieee.org)}
 \thanks{The research leading to these results has received funding from the European Commission's Horizon 2020 Framework Programme (H2020/2014-2020) under grant agreement No. 646470, SmarterEMC2 Project.}
}

\maketitle

\IEEEpeerreviewmaketitle

\begin{abstract}
Wireless power transfer (WPT) technologies have been widely used in many areas, e.g., the charging of electric toothbrush, mobile phones, and electric vehicles.  This paper introduces fundamental principles of three WPT technologies, i.e., inductive coupling-based WPT, magnetic resonant coupling-based WPT, and electromagnetic radiation-based WPT, together with discussions of their strengths and weaknesses. Main research themes are then presented, i.e., improving the transmission efficiency and distance, and designing multiple transmitters/receivers. The state-of-the-art techniques are reviewed and categorised.  Several WPT applications are described. Open research challenges are then presented with a brief discussion of potential roadmap. 
\end{abstract}


\section{Introduction}
\IEEEPARstart{H}{ave} you ever imagined that, one day in the future, all electricity pylons will disappear? Wireless power transfer (WPT) is an innovative technology that may make this a reality by revolutionizing the way how energy is transferred and finally makes our lives truly wireless. 

WPT technologies can be dated back to early 20th century. Nikola Tesla, a pioneering electronic engineer, invented the Tesla Coil aiming to produce radial electromagnetic waves with about 8 Hz frequency transmitted between the earth and its ionosphere, thereby transferring energy \cite{1}. Nowadays,  are widely used in daily life for charging mobile devices, e.g., cell phones, laptops, tablets. In addition, WPT technologies have been proved useful in wireless sensor networks \cite{3,4}, whose lifetime can be extended \cite{41, 5}. In 2009, Sony's research and development department in Japan has made a breakthrough in developing highly efficient WPT-based TVs. In 2014, Apple inc. released iWatch with wireless charging design. 

Generally speaking, the WPT technologies can be categorised into inductive coupling-based WPT, magnetic resonant coupling-based WPT, and electromagnetic radiation-based WPT. The inductive coupling-based WPT makes use of magnetic field induction theory and belongs to near-field transmission technique. The magnetic resonant coupling-based WPT relies on the same resonant frequency at both receiver side and transmitter side, enabling longer distance power transmission. The electromagnetic radiation-based WPT has longest transmission distance by using microwave frequencies. Every WPT technology has its own advantages and disadvantages. The main research themes of all WPT technologies focus on improving the transmission efficiency and distance. In addition, driven by industrial needs, multiple transmitters/receivers design has attracted much attention. However, several open research challenges exist, hindering the public acceptance of WPT technologies. Hence, it is crucial to understand and address these challenges for paving the way for a brighter future of the WPT applications. 

This paper provides a survey of the state-of-the-art WPT technologies and then presents open research challenges and potential roadmap. In the remainder of this paper, we first describe fundamental principles of WPT technologies in Section~\ref{section2}, categorise main research themes of WPT technologies in Section~\ref{section3}, and then discuss the applications of WPT technologies in Section~\ref{section4}. Open research challenges of WPT are subsequently identified in Section~\ref{section5}, together with a brief discussion of promising roadmap.

\section{Fundamental Principles}
\label{section2}

There are three categories of WPT technologies: inductive coupling, magnetic resonant coupling, and electromagnetic radiation.

\subsection{Inductive Coupling Approach}
\begin{figure}[!t]
		\begin{center}
		\includegraphics [width=3.2in]{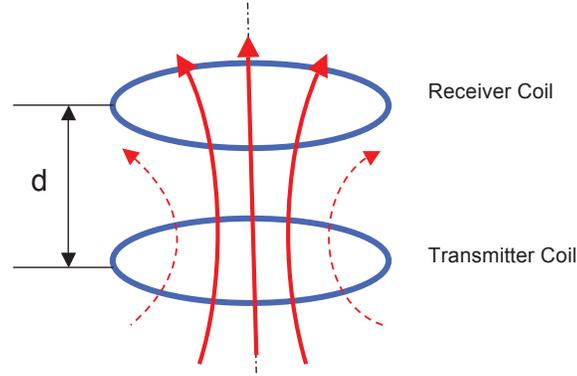}
		\end{center}
		\caption{Scheme of Inductive Coupling-based Wireless Power Transfer.}
		\label{DIW}
\end{figure}

Inductive coupling-based WPT makes use of magnetic field induction as shown in Fig.\ref{DIW}.  It is a typical near-field transmission technique: the state-of-the-art technologies can support transmission distance of mm or cm \cite{2}. 
The fundamental theories of inductive coupling-based WPT include the use of Biot-Savart's Law and Faraday's Law. The Biot-Savart's Law is used for calculating the magnetic field produced by an arbitrary current distribution:
 \begin{equation}
 {\bf B}=\frac{\mu_{0}}{4\pi}\oint\frac{{\bf I}d_{{\bf I}}\times{{\bf r}}}{|{\bf r}|^{3}}
 \end{equation}
where $\mu_{0}$ is the magnetic constant, ${\bf I}$ is the current in the transmitter coil, $d_{{\bf I}}$ is a vector whose magnitude is the length of the differential element of the wire, and ${\bf r}$ is the full displacement vector from the wire element to the point at which the field is being computed.
The Faraday's Law can then be used for calculating the induced voltage $V_{Ind}$ over the receiver coil as the rate of magnetic field ${\bf B}$ change through an effective surface area $S$ by
  \begin{equation}
  V_{Ind}=-\frac{\partial}{\partial t}\oint{{\bf B}} \cdot d_{s}.
  \end{equation}

One significant drawback of the inductive coupling-based WPT is its short transmission distance. Moreover, when the transmitter coil and the receiver coil are not well aligned, the power transmission efficiency (PTE) drops significantly.
Despite of these weaknesses, inductive coupling-based WPT is often advantageous with respect to its simple design and high safety, therefore has been broadly used in many applications including the charging of toothbrush, laptops, mobile phones, and medical implants \cite{1}.
 
\subsection{Magnetic Resonant Coupling Approach}

\begin{figure}[!t]
	\begin{center}
		\includegraphics [width=3.3in]{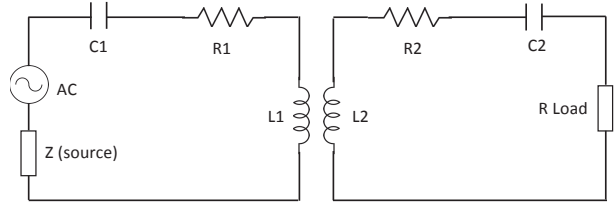}
	\end{center}
\caption{The equivalent circuit of magnetic resonant coupling-based wireless power transfer where $ L1$, $ L2$, $C1$, $C2$, $ R1$ and $R2$ are the inductance, the capacitance, and resistance of the transmitter and the receiver coils, respectively, and $RLoad $ is the load resistance. }
\label{MRC}
\end{figure}

 A typical magnetic resonant coupling system consists of two electromagnetic subsystems with the same natural resonance frequency, enabling efficient power transfer \cite{8}. Such a system can be represented by a standard RLC circuit consisting of a resistor, an inductor, and a capacitor as shown in Fig.~\ref{MRC}.  When the transmitter coil is excited by the source, the transmitter and the receiver coils are magnetically coupled. There are two key factors that determine its PTE:   
\begin{enumerate}
\item The Q-factors of the resonators (Q)   
\item The mutual coupling strength (M) \cite{30,46}.
\end{enumerate}
 
The Q factor is often defined by \cite{8}:
\begin{equation}
{Q}=\frac{1}{R}\sqrt{\frac{L}{C}}=\frac{\omega_{0}L}{R}
\end{equation}
where $\omega_{0}=\frac{1}{\sqrt{LC}} $ denotes the resonant frequency. A higher Q indicates less energy loss. 

 Mutual inductance can be written as \cite{8}:
\begin{equation}
{M}=k\sqrt{L_{1}*L_{2}}
\end{equation}
where $k$ is the coupling coefficient determined by the distance between transmitter and receiver coils. For circle coils, it can be written as \cite{8}
\begin{equation}
{k}=\frac{1}{[1+2^{\frac{2}{3}}{(\frac{h}{\sqrt{r_{t}*r_{r}}})^{2}}]^{\frac{3}{2}}}
\end{equation} 
where $h$ denotes the distance between two coils, and $r_t$ and $r_r$ are the radius of the transmitter and the receiver coils.

Compared with the inductive coupling approach, the magnetic resonant coupling approach can transfer power to a longer distance. Furthermore, it is non-radiative, so that it does not require line of sight and has almost no harm to human~\cite{1}. However, similar to the inductive coupling approach, the magnetic coupling approach is also sensitive to misalignment. Another problem is that it is difficult to adjust the resonance frequency when charging multiple devices \cite{7}. 

 \subsection{Electromagnetic Radiation Approach}
Improving PTE is always an important issue  in the development of WPT technology. Strong directional and energy concentrated laser beam, which has similar properties as microwave beam, is worth being investigated.

A typical electromagnetic radiation approach requires a microwave source, a waveguide, a transmitting antenna, and a receiving antenna. 
The energy is initially transmitted from the microwave source to the waveguide. The transmitting antenna then emits electromagnetic wave uniformly. The receiving antenna adopts silicon-controlled rectifier diode to collect microwave energy and convert it into direct current.

The main advantage of electromagnetic radiation approach is its high PTE over long distance. But it often has radiation concerns and requires the line of sight transmission. Table \ref{table1} shows a summary of advantages and disadvantages of different wireless power transfer technologies. 

\begin{table*}[!t]
\caption{Summary of advantages and disadvantages of different wireless power transfer technologies}
\centering 
\begin{tabular}{l c c} 
\hline \hline
&&\\
WPT Technology Category & Advantages & Disadvantages \\
&&\\
\hline 
&&\\
Inductive Coupling & Sample, safe, and high transfer  & Short transmission distance, \\
 & efficiency in short distance & needs accurate alignment \\
&&\\
\hline
&&\\
Magnetic Resonant Coupling & Long transmission distance, & Difficult to adjust resonant  \\
 &  no radiation & frequency for multiple devices \\
&&\\
\hline
&&\\
 Electromagnetic Radiation & Very high transmission  & Radiation,  \\
 & efficiency over long distance & needs line of sight \\
 &&\\
\hline \hline
\end{tabular}
\label{table1}
\end{table*}

\section{Main Research Themes}
\label{section3}
Although WPT technologies have been successfully utilized to develop a number of commercial products, there are still many areas of improvement. Reviewing the literature, the main research themes can be summarized as:  
\begin{enumerate}
\item Improving the transmission efficiency and distance, 
\item Multiple transmitters/receivers design.
\end{enumerate}

\subsection{Improving the transmission efficiency and distance }
\label{section3.1}

There are several impact factors affecting the PTE and the transmission distance. These impact factors are summarized in Table~\ref{2}. 
\begin{table}[!t]
\centering
\caption{The summary of impact factors}
\label{2}
\begin{tabular}{l l}
		\hline \hline
		Impact Factor & Description [reference] \\ \hline
		Coil Design & Coil material \cite{9, 10, 11} \\
		 & Coil size \cite{12, 14} \\ 
		 & Coil geometry \cite{14, 13} \\  \hline
		Coil Alignment & Lateral misalignment \cite{20, 6} \\ 
		&  Angle misalignment \cite{6} \\ \hline
		Circuit Design &  Frequency \cite{15,12} \\ 
		& Compensating circuit \cite{16} \\ \hline
		Environmental Analysis & Temperature, humidity \cite{8, 12}\\ 
		&
		Ferrite Core\\ \hline
		Multiple Transmitters & Add additional coils \cite{17}.\\ \hline \hline
	\end{tabular}
\end{table}

\subsubsection{Coil Design}

Optimising the coil material can improve the system's PTE. In \cite{9}, it has been shown that using a high Q planar-Litz coil can get almost 20 \% efficiency improvement. In \cite{10}, a magnetoplated wire (a copper wire) has been used, and in \cite{11}, multi-layer, multi-turn tubular coils have been adopted for improving the transmission efficiency, with a study of the skin-effect.
		
In the inductive coupling-based WPT, the ratio of the transmitter coil diameter and transmit distance (air gap) can also affect the transmission efficiency. Assuming a constant Q-factor, if the air gap ($L$) is smaller than the half of the coil diameter ($A$),  i.e., $L/A<0.5$, the transmission efficiency will become greater than 80\%. Furthermore, if $L/A<0.25$, the transmission efficiency can reach $90\%$ \cite{12}.

The coil geometry can also influence the PTE. Spiral coil, flat coil, square coil, and circular coil have different PTEs according to the studies in \cite{13}. In \cite{14}, asymmetric coupling resonators have been proposed that consist of three different size transmitter coils and two different size receiver coils. 

\subsubsection{Coil Alignment}

One of the important factors affecting the PTE is the coil misalignment between the transmitter coil and the receiver coil \cite{13, 6, 18,19}. There are two forms of misalignment \cite{6,18,45}

\begin{itemize}
\item Lateral misalignment: when the transmitter coil and the receiver coil are placed in parallel, their centres do not only have horizontal distance $\Delta$, but also have the vertical distance $d$. 
\item Angular misalignment: the receiver coil is turned by an angle $\vartheta$ when the centres of the transmitter coil and the receiver coil are well aligned.
\end{itemize}
The magnetic resonance coupling WPT system with lateral misalignment has been studied in \cite{20} that uses ANSYS model to simulate the relation between the transfer efficiency and the lateral misalignment distance. The high efficiency misalignment range can then be determined. In practice, the coils' lateral and angular misalignment may happen together. However, most of existing work tends to analyse the lateral misalignment and the angular misalignment, respectively. Therefore, there exists a research gap here. 

\subsubsection{Circuit Design}
To maintain a high PTE, we can find the optimal operating frequency, e.g., by frequency tracking. In \cite{15}, an automatic frequency turning method has been proposed to find the possible maximum PTE. However, due to the skin effect and the proximity effect increasing winding resistances, the operating frequency shall not be too high \cite{12}. In \cite{16}, a compensating system has been proposed to improve the PTE by using a serious-shunt mixed-resonant circuit, resulting in a high transmission efficiency over long distance. 

\subsubsection{The Environmental Analysis}

The temperature and humidity \cite{8} are important factors that affect the PTE. Some semiconductors are also sensitive to the temperature \cite{12}. In addition, adding a ferrite core can improve the magnetic field strength and increasing magnetic flux, thus leading to higher PTE.  Adding a third coil may also improve the PTE. In \cite{17}, an optimal design has been proposed that makes use of an intermediate coil between the primary coil and the secondary coil. It can boost the effective self-inductance and magnetizing inductance in the primary coil, thereby increasing the apparent coupling coefficient.  

\subsection{Multiple transmitters/receivers design}

Exploring the use of multiple transmitters and receivers is another important research theme. This is because the ability of transferring power from a single transmitter to multiple receivers is often required in practice \cite{22}. Moreover, the transmission efficiency varies as the number of the transmitters/receivers changes. 

In \cite{22}, a method has been proposed to analyse the influence of multiple transmitters or multiple receivers in the WPT system. Compared with the single transmitter/receiver coil case, it has been found that increasing the receiver coils can improve the transmission efficiency. In contrast, increasing the number of transmitter coils could degrade the transmission efficiency if the coupling coefficient from a newly added transmitter to the receiver is small.

Matching the resonant frequency in multiple transmitters/receivers in the magnetic resonant coupling-based WPT system has been found difficult \cite{7}. An algorithm has been studied to determine the resonant frequency in a 2 by 2 array multiple transmitters/receivers WPT system. The PTE can be maintained at certain high level without changing the system operating frequency. In \cite{28}, a multi-receiver WPT system  has been presented that analyses both the PTE and the power distribution among the receivers. The power distribution depends on both the load impedances and the relative position of the receivers to the transmitter. In addition, it has been found that using a repeater can extend the transmission distance.

\section{Applications}
\label{section4}
 
 \subsection{Home Electronics:}
Besides the charging of smart phones, cameras, and iWatches, the WPT technologies can be used for charging TVs.
In \cite{24}, a 150 watt wireless charging circuit has been presented that makes use of three self-resonators for charging 47 inch LED TV. The transmitter resonator and the intermediate resonator are placed in vertical, whilst the receiver resonator and the intermediate resonator are placed in parallel. The PTE has been shown as up to 80\% when the frequency is 250MHz. 
 
 \subsection{Medical Implants:}
One important application of WPT technologies is the medical implants, e.g., cardiac implant \cite{25}. Most medical implants only need tiny voltage, and they will not use external charging once transplanted into the human body.  In many medical applications, a WPT transmitter can be mounted outside of the human body whereas the receiver which is implanted into the human body. In this setup, the surgeries of replacing batteries can be avoided.

 \subsection{Electric Vehicles:}

In \cite{26}, the concept of the roadway-powered electric vehicle has been proposed which can achieve 100 kW power output with $80\%$ PTE at the 26 cm transmitter/receiver distance. In order to compensate the reactive power, a series-compensation system has been adopted that is proved with high PTE and high tolerance of lateral misalignment. Such a system consists of two main parts: transmitter part with an inverter and the power line, and the receiver part with pick-up modules, rectifiers, and regulators.

 \subsection{Wireless `Grid':} 
Electrical grid is an essential part of human life due to its capability of delivering electricity from suppliers to consumers.  WPT-based `grid' would eliminate the need of using wire transmission grids to deliver electricity \cite{44}. In the future, it  may be possible to transfer power from remote renewable energy sources, such as wind farms and solar array fields, to consumers. In addition, it could enable the collection and utilization of micro-power from ambient sources - a technique known as power harvesting \cite{42,43}.

\section{Open Research Challenges and Roadmap}
\label{section5}
 
\subsection{Self-adaptation to Misalignment}

In practice, lateral misalignment and angular misalignment will inevitably happen that reduce the PTE and the transmission distance. One potential approach of mitigating misalignment is to track the power output on the load at the receiver side and then adjust receiver parameters to maximise the power output. However, this may involve multiple parameters adaptation such as lateral misalignment, angular misalignment, and operating frequency. The research challenge is on how to adjust these parameters in a real-time manner whist guaranteeing the global optimal power output. Alternatively, we could adapt the transmitter parameters by obtaining feedback from the receiver side such as building a feedback link by using Zigbee modules. The Zigee technology is based on IEEE 802.15.4 protocol and used in low cost, low power consumption, low complexity, and short distance networking applications. The future research challenges lie on the mutual self-adaptation of both transmitter and receiver's parameters in order to obtain maximum PTE. 
 
\subsection{Self-adaptation to Coil Length and Geomery}
As aforesaid in Section~\ref{section3.1}, both coil length and coil geometry have effects on the PTE and the transmission distance. With the same coil length, the coil geometry can be optimised, and vice versa. In this context, it is crucial that the transmitter coil and/or the receiver coil can adaptively change their length/geometry to improve the PTE. To do that, it is essential to build theoretical models relating the coil length/geometry with the PTE. Moreover, the coil itself shall be flexible enough enabling us to change its length and geometry. Therefore, the modelling of PTE as a function of coil length and geometry, and the coil flexible design are of prime research interests in the future. 
 
\subsection{Small Animal Protection}
WPT systems may potentially cause healthy issues of small animals. To protect small animals from the WPT system, it will be necessary to explore small animal detection system. For example, when the electric vehicle is being charged, the small animal (such as dog or cat) may stay under the vehicle, i.e., between the transmitter coil and the receiver coil, that  may harm the small animal. Therefore, a suitable small animal protection system will be needed to detect them, e.g., turn off the charging system when appropriate.

\section{Conclusion} 
\label{section6}

WPT technologies can be divided into three types: inductive coupling-based WPT, magnetic resonant coupling-based WPT, and electromagnetic radiation-based WPT. The first two WPT technologies belong to near-field transmission techniques, and the last one is a far-field transmission technique. They present different features and can be used in different scenarios. Although the principles of these three types WPT are different, the common feature is the lower PTE compared with the wired power transfer. The state-of-the-art research focuses on improving transmission efficiency and distance, and enabling multiple devices charging. The challenges issues lie on the adaptation of system (either transmitter or receiver) parameters, such as coil length and geometry, to address the misalignment problems since the misalignment often occurs in practice.

\bibliographystyle{IEEEtran}
\bibliography{IEEEabrv,reference}
\end{document}